# Optimal Algorithmic Monetary Policy:
# Blockchain for Digital Money

Luyao Zhang[1] and Yulin Liu[2]

10/6/2021

Centralized monetary policy, leading to persistent inflation, is often inconsistent, untrustworthy, and unpredictable. Algorithmic stablecoins enabled by blockchain technology are promising in solving this problem. Algorithmic stablecoins utilize a monetary policy that is entirely rule-based. However, there is little understanding of how to optimize the rule. We propose a model that trade-off the price for supply stability. We further study the comparative statics by varying several design features. Finally, we discuss the empirical implications for designing stablecoins by the private sector and Central Bank Digital Currency (CBDC) by the public sector. (JEL: G11, G12, G17, C55)

Keywords: algorithmic monetary policy, stablecoins, CBDC, blockchain, rule vs discretion

[1]Zhang: Data Science Research Center and Social Science Division, Duke Kunshan University, Suzhou, China 215316; [2]Liu: Bochsler Finance & Associés SA, Gotthardstrasse 26, 6300 Zug, Switzerland. Each author contributed equally to this research. Corresponds to Zhang (lz183@duke.edu) and Liu (yulin@bochslerfinance.com). Acknowledgments. We thank Prof. Campell Harvey for his insightful comments.

## 1 Introduction

Kydland and Prescott (1977) advocate rules rather than discretion in policy implementation. In terms of monetary policy, discretionary power results in higher inflation without reducing unemployment due to rational expectations. However, there is no guarantee that a centralized policymaker will commit to the rules of maintaining price stability. Algorithmic stablecoins enabled by blockchain technology are promising in solving this problem.

Stablecoins are one type of decentralized finance applications (Harvey, Ramachandran, and Santoro 2021) intended to remedy cryptocurrencies' excess volatility. Figure 0 represents three types of stablecoins and their typical example.[1] Among them, the algorithmic stablecoin, implements a monetary policy that is entirely rule-based. The rule is programmed in a smart contract on the Ethereum blockchain (Buterin, 2013), which is tamper-proof in general (Mohanta, Panda, and Jena 2018). The algorithmic stablecoins automatically and proportionally adjust supply in the wallet in response to demand shocks. For example, when the price appreciates above its target, the supply in everyone's wallet increases until the price goes back to the target. Vice versa, when the price depreciates below the target, everyone's holding of the stablecoin decreases. As the adjustment of token number is proportional, users' percentage of token stays unchanged. Thus, algorithmic stablecoins inherit the best from both conventional fiat currency and bitcoins alike cryptocurrencies. On the one hand, like traditional fiat currency, it has contingent scarcity so that supply can adjust to match demand[2]; on the other hand, like other cryptocurrencies, it is open-source and decentralized.

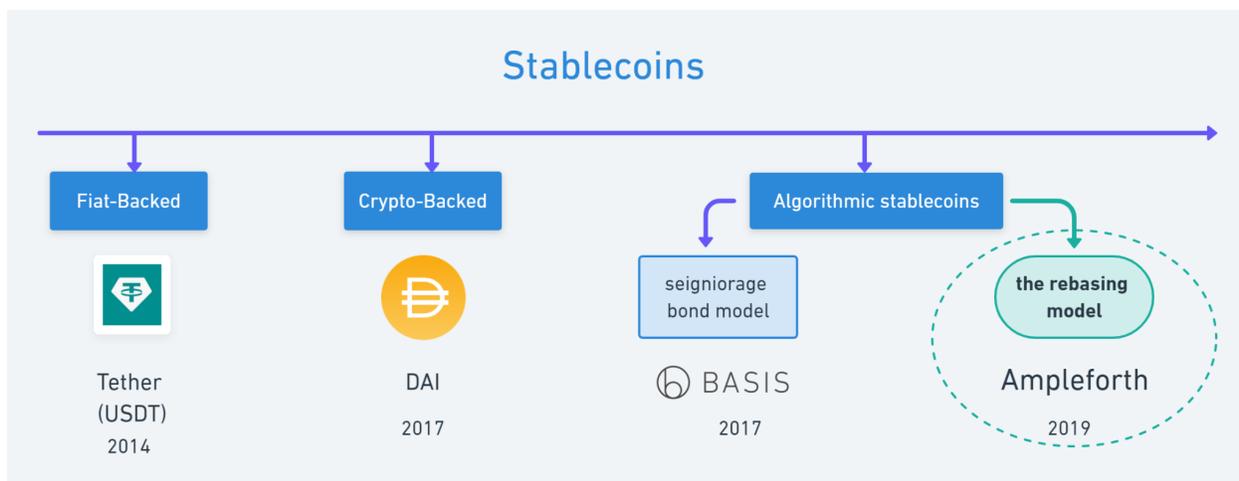

Figure 0: Types of Stablecoins

The theoretical concept of stablecoins has appeared as Hayek money. Hayek (1990) advocates a system of private currencies that financial institutes compete to provide a currency of the most excellent stability. However, the recent innovative implementation is the algorithmic stablecoins. For monetary expansion, money is created out of algorithms and distributed to users according to token share. Unlike the quantitative easing that redistributes wealth, the expansion does not dilute any user's stake of supply. For monetary contraction, a fraction of money disappears when the token price is below the target. Unlike the negative interest rate where depositors could withdraw their money from banks, there is no physical form of the algorithmic stablecoins for users to hide from the contraction.

---

[1] We provide a comprehensive discussion for the different types in Section 1.1.
[2] Different from algorithmic stable coins, bitcoin-alike cryptocurrencies are synthetic commodities (George 2015) with absolute scarcity.

However, there is little understanding about how to optimize the rule set in the smart contract. In Section 2, we propose a model that trade-offs between the price and supply stability. In Section 3, we further study the comparative statistics by varying several design features. Finally, in Section 4, we discuss the empirical implications and further research for industry applications. In the subsection below, we discuss the background and literature.

1.1 background and literature

Due to the hefty price volatility, Bitcoin fails to deliver its vision of a "peer-to-peer electronic cash system". Instead, Bitcoin is adopted more as an investment vehicle and a store of value (Liu and Zhang 2021). Unlike Bitcoin and other cryptocurrencies with large price swings, stable coins peg their values to sovereign currencies, typically the US Dollar. As of June 2021, the total market cap of stable coins attained $112 billion. Stable coins are heavily used for daily transactions and as a crypto safe-haven against other volatile crypto assets (Baur and Hoang 2020; Wang, Ma, and Wu 2020). The majority of the stablecoins are fiat-backed. Namely, every stablecoin in circulation is collateralized by one US Dollar in a centralized custodian account. As in Figure 0, Tether (USDT)[3] is one of this type. The major problem of the fiat-backed stable coins is that a single entity controls the collateral, exposing the stable coins to centralization risks. Lyons and Viswanath-Natraj (2019) analyzes the price stability of fiat-backed stablecoins.

Crypto asset-backed stable coins solve this problem by replacing the fiat collateral with the crypto asset, such as Ethers. One typical example is Dai.[4] Users could open a collateralized debt position by pledging ethers as collateral in a smart contract. Users could then borrow Dai tokens against her collateral. Due to the vast volatility of ether price, the collateral ratio of Dai tokens is set at 150% to maintain stability. Over-collateralization leads to capital inefficiency, i.e., minting a stablecoin of $X requires at least a collateral of $1.5X. Li and Mayer (2021) developed a comprehensive dynamic model to address the optimal design of the crypto asset-backed stable coins.

Algorithmic stablecoins were born against this backdrop. An automated algorithm adjusts the supply to maintain price stability. There are two types of algorithmic stable coins: the seigniorage bond model (d'Avernas, Bourany, Vandeweyer, 2021) and the rebasing model. The former has a dual-token or triple-token system. Basis[5] is one representative. When the stablecoin price falls below its target, the system contracts coin supply by minting new bonds in exchange for the stablecoins. To incentivize stable coin holders to swap coins for bonds, the bond price is set at lower than the target value. The purchased stablecoins are burned, and thus the money supply decreases. When the stablecoin price rises above the target, the system expands coin supply by minting new stablecoins and selling them to bondholders at the target value. By adjusting the supply of the stable coin to counteract the fluctuating demand, the system achieves price parity.

---

[3] https://tether.to/
[4] https://makerdao.com/
[5] https://www.basis.io/

The seigniorage bond model has a vital issue. When users expect a long contraction period in the crypto winter, the system has to lower the bond price to purchase sufficient stable coins. The bond price could hit the price floor in certain circumstances, i.e., $0, but the stablecoin price is still below parity. In such cases, the system falls into the death spiral.

The second type of algorithmic stable coin does not rely on a bond token for supply adjustments. Instead, it simply reduces or increases the number of stablecoins in every wallet. Ampleforth[6] is one typical example. It is like successfully exercising a negative or positive interest rate to every bank account. In traditional finance, bank users could withdraw money from their accounts and hold cash to avoid the negative interest rate. However, in the crypto world, there is no physical form of the stable coin to withdraw and to hide from algorithmic adjustment. We research the optimal algorithmic monetary policy for the rebasing model.

Besides advising the design of stablecoins by the private sectors, our research also has implications for the Central Bank Digital Currency (CBDC) issued by the public sector. CBDC is a digital form of legal tender that has been widely studied and discussed among academics and policymakers (Agur et al. 2021). CBDC is a new digital form of money issued by central banks and could provide a certain extent of anonymity. Unlike central bank reserves, available only to wholesale banks for interbank settlements, CBDC is issued for all bank and non-bank entities. Assenmacher et al. (2021) build a general equilibrium search model to study the impact of the CBDC on social welfare and find that CBDC dis-intermediates banks and improves resource allocation. Barrdear and Kumhof (2019) set up a dynamic stochastic general equilibrium (DSGE) model to study the impact of CBDC on economic output and find that implementing CBDC could significantly improve the stabilization of business cycles and raise GDP by up to 3%. Most existing literature addresses CBDC with a discretionary monetary policy.[7] However, a discretionary CBDC is still not a solution for mistrust in policy makers. Fortunately, the central banks can also settle CBDC transactions via Distributed Ledger Technology (DLT) in a permissioned blockchain. A rule-based CBDC on blockchain can further boost trust in policy makers. Our paper paves the way for designing an algorithmic CBDC.

## 2 The model

---

[6] https://www.ampleforth.org/

[7] Among the pioneering literature, Schilling, Villaverde, and Uhlig (2020) demonstrate the CBDC impossible trinity: economic efficiency, financial stability, and price stability; Brunnermeier and Niepelt (2019) provide equivalence between private money such as cryptocurrencies and public money such as CBDC and their effect on financial stability; Niepelt (2019) identify conditions under which the introduction of CBDC could change the macro outcome or not.

## 1.1 The objective function

An algorithmic stablecoin should achieve stability in both price and purchasing power to be an ideal decentralized payment system. (Mita et al. 2019). This objective is a trade-off between the deviation of price from its target and the volatility in supply. Denote by $P_t$ and $S_t$ the price and supply of the stable coin at time $t$. Considering a finite period of $t = 1, 2, ..., n$, the objective is thus to minimize the following loss function:

$$E\{\sum_{i=1}^{n} dP_t^2 + \lambda dS_t^2\};$$

$dP_t = (P_t - P^*)/P^*$, where $P^*$ is the target price, $dS_t = (S_t - S_{t-1})/S_{t-1}$, and $\lambda$ is the relative weight for the trade-off between price and supply stability.

## 1.2 The policy parameters

As the frequent adjustment of the number of tokens in the wallet causes inconvenience to users, algorithmic stable coins set an inactive price range, denoted by $[P^* - A, P^* + A]$. When price $P_t$ falls within this range, the supply remains unchanged, i.e. $S_{t+1} = S_t$; when $dP_t = P_t - P^* > A$, supply increases such that $dS_{t+1} = S_{t+1} - S_t = dP_t/B$; likewise, when $dP_t < -A$, supply decreases such that $dS_{t+1} = -dP_t/B$. $A$ and $B$ are hence the policy parameters. A policymaker chooses the optimal $A$ and $B$ to minimize the objective function.

## 3 The simulations

We assume that the market cap, $Y_t = P_t \times S_t$, following the standard Geometric Brownian Motion (Ross 2014):

$Y_t = Y_0 e^{(u - \sigma^2/2)t + \sigma W_t}$

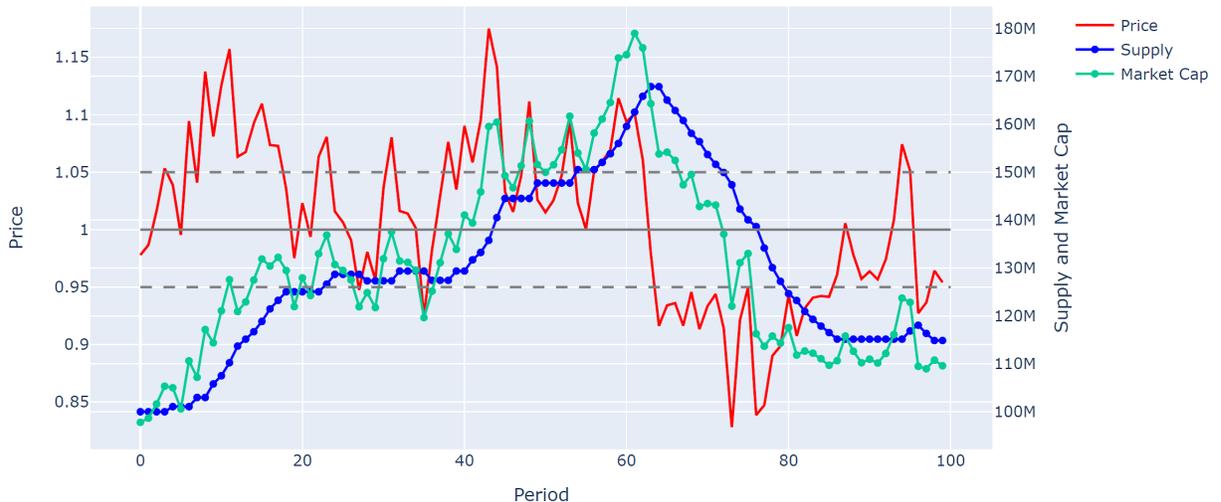

Figure 1: Price, Supply, and Market Cap.

Figure 1 simulates the market cap in GBM with the volatility $\sigma = 0.05$, the drift $\mu = 0$, and the initial market cap $Y_0 = 100$ million USD (M) for a period of $n = 100$; the supply policy is simulated with $A = 0.05$, $B = 5$, and $P^* = 1$, i.e., the supply remains unchanged when the token price is within the range of $[0.95, 1.05]$ and adjustment at a rate of $dS = dP/5$ otherwise. When the **market cap** changes abruptly, the token **price** could run out of the inactive range suddenly, and then the **supply** adjusts accordingly to bring the token price back to the inactive range.

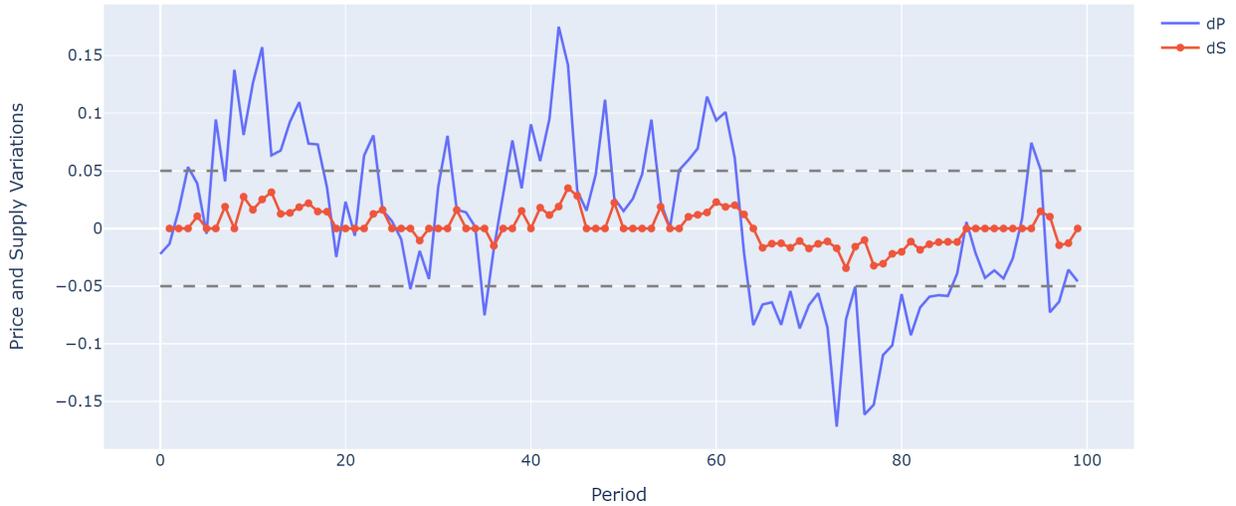

Figure 2: the deviation of price from its target and the supply volatility.

Figure 2 shows the deviation of price from its targets $dP$ and the supply volatility $dS$. For $\lambda = 1$, the price deviation and the supply volatility have equal weights. In the figure the price deviations are obviously larger than the supply volatility, implying the policy parameter $A = 0.05$ and $B = 5$ are apparently not optimal: a slight decrease in B is an obviously better alternative since it increases the price stability on a much larger scale than decreases the supply stability. In the next session, we simulate the comparative statistics for the loss function and the optimal policy.

## 4 The result of comparative statics

We have three main results in simulating the comparative statics: (1) ceteris paribus, a more extensive inactive range (a larger A) results in a greater loss. (2) ceteris paribus, we put more weights on the supply volatility, a larger $\lambda$, the optimal policy adjusts supply less intensively, i.e. a larger smaller $A$ and $B$. (3) the results of (1) and (2) are robust when varying the volatility, i.e. $\sigma$, and the drift, i.e. $\mu$, of the market cap.

### 4.1 Comparative statics for the loss function

We start with $\lambda = 1$ as the benchmark. Figure 3 illustrates the comparative statics for the loss function when varying policy parameters [    ] and $B = [1, 10]$. We simulate the result ten times for each parameter pair and take the average loss as an estimator of the expected loss. The result shows that in general, ceteris paribus, the loss increases with $A$, i.e., a wider inactive range leads to a greater loss.

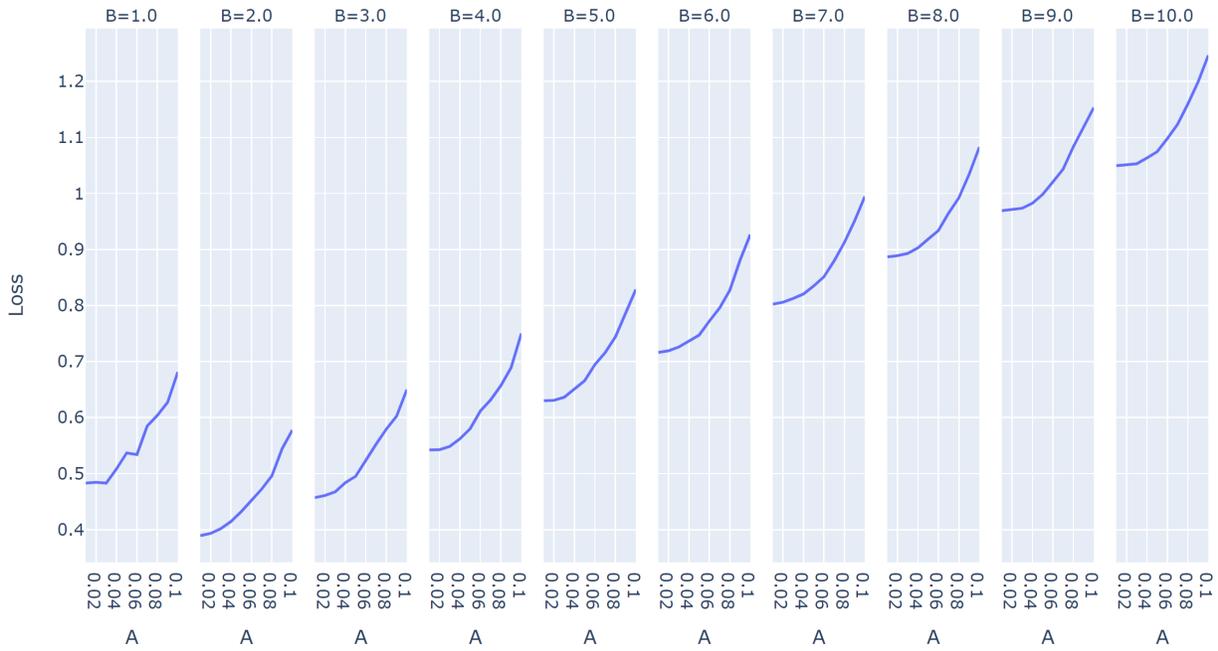

Figure 3: Comparative Statics for the Loss Function

4.2 Comparative statics for the optimal policy

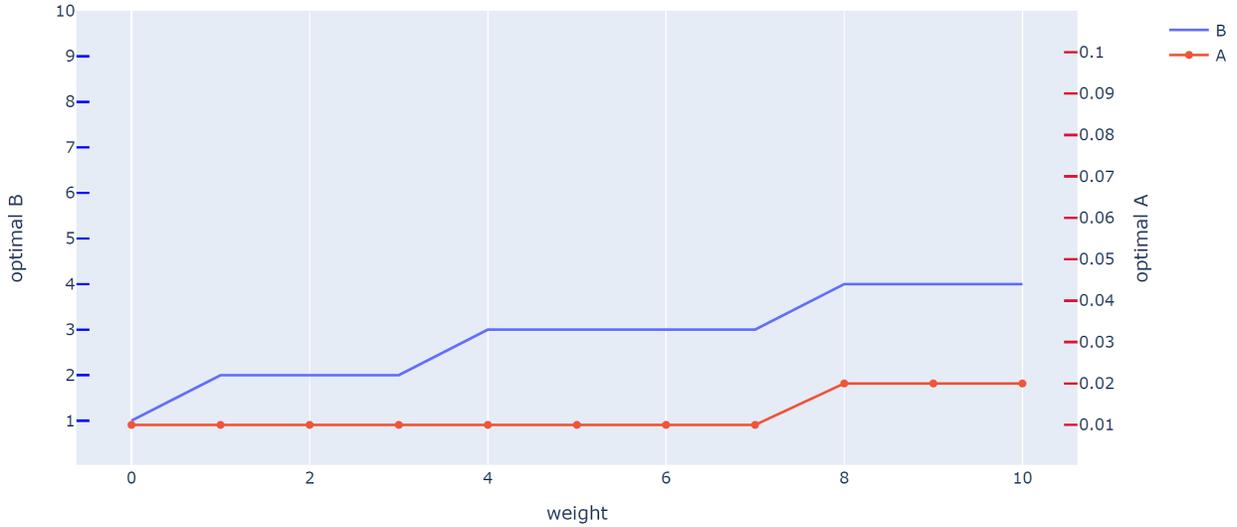

Figure 4: Comparative Statics for the Optimal Policy

Figure 4 illustrates the comparative statics for the optimal policy when varying weights of the loss function, i.e. $\lambda = [0, 10]$. Parameter A defines how frequently the supply adjustment takes place. Larger values of A lead to less frequent adjustment. In contrast, parameter B defines how radical the adjustment is. Greater values of B indicate smoother supply adjustments. The result shows that when we have a higher weight of the supply stability, the optimal B increases to balance the loss in price deviation and supply variation.

4.3 Robustness

Ceteris paribus, Figures 5 and 6 show the robustness result for (1) and (2) with a positive drift $\mu = 0.01$ and a negative drift $\mu = -0.01$ respectively.

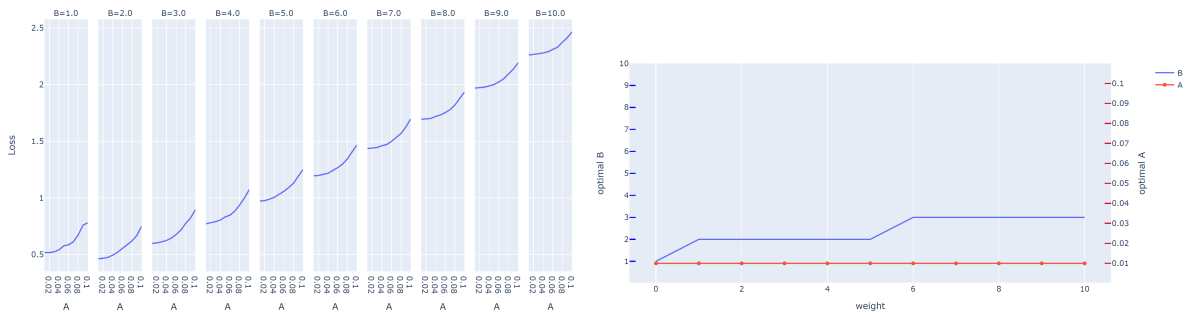

Figure 5: Robustness with a positive drift

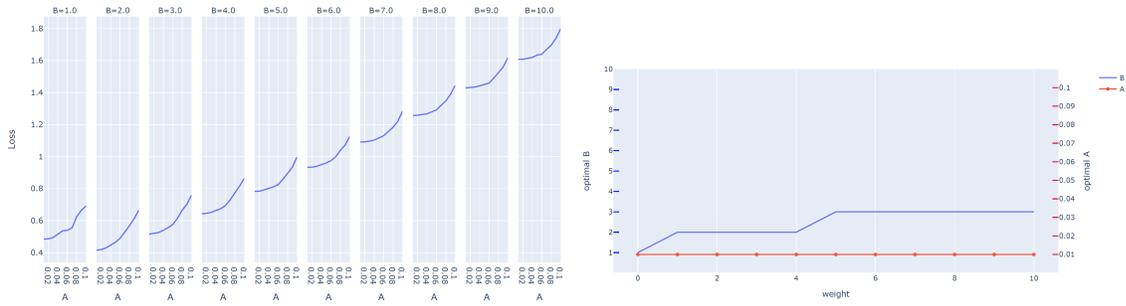

Figure 6: Robustness with a negative drift

Ceteris paribus, Figure 7 shows the robustness for (1) and (2) with a larger volatility of the market cap, i.e. $\sigma = 0.5$.

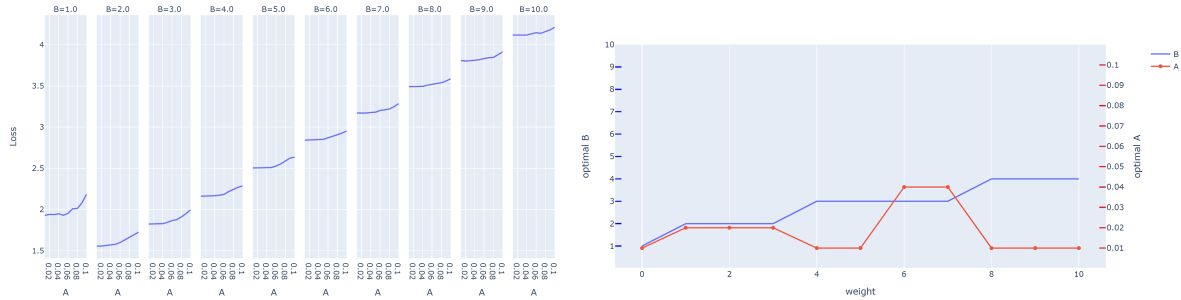

Figure 7: Robustness with a larger volatility

## 5  Conclusion and further research

As a cryptocurrency, stablecoins are designed to be decentralized. Moreover, stablecoins can serve as a safe haven against the unstable coins such as bitcoins and ethers (Baur and Hoang 2021, Wang, Ma, and Wu 2020). Among stablecoins, algorithmic stablecoins allow automated and rule-based monetary policy. The real-world applications include Ampleforth (AMPL)[8], Empty Set Dollar[9], and Base Protocol[10], etc. Using the rebase history data of AMPL,[11] we plot the historical price, supply, and market cap in Figure 8. We further visualize the deviation of price from its target and the supply volatility in Figure 9.

Figures 8 and 9 demonstrate that the market overvalued AMPL at the genesis, and the correction overshoot in the following three months. It took around half a year for the price to stabilize at $1. In mid-2020, the start of the DeFi craze, due to the inflow of hot money in the crypto space, AMPL price stayed consistently above parity, leading to the aggressive expansion of the money supply.

---

[8] https://www.ampleforth.org/
[9] https://www.emptyset.finance/
[10] https://www.baseprotocol.org/
[11] https://www.coin-tools.com/ampl/ampl-rebase-history/

Figure 9 shows that the price deviations are obviously larger than the supply volatility, implying the current policy parameter[12] in AMPL are apparently not optimal for a equal weights, i.e. $\lambda = 1$: a slight decrease in B is an obviously better alternative since it increases the price stability on a much larger scale than decreases the supply stability. Our approach can thus be applied to advise Algorithmic Stablecoin design.

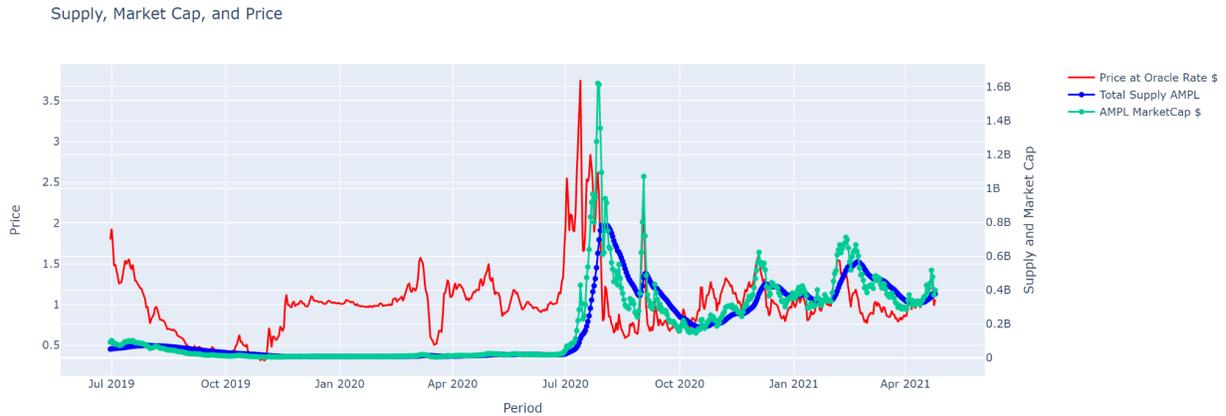

Figure 8: Price, Supply, and MarketCap of AMPL

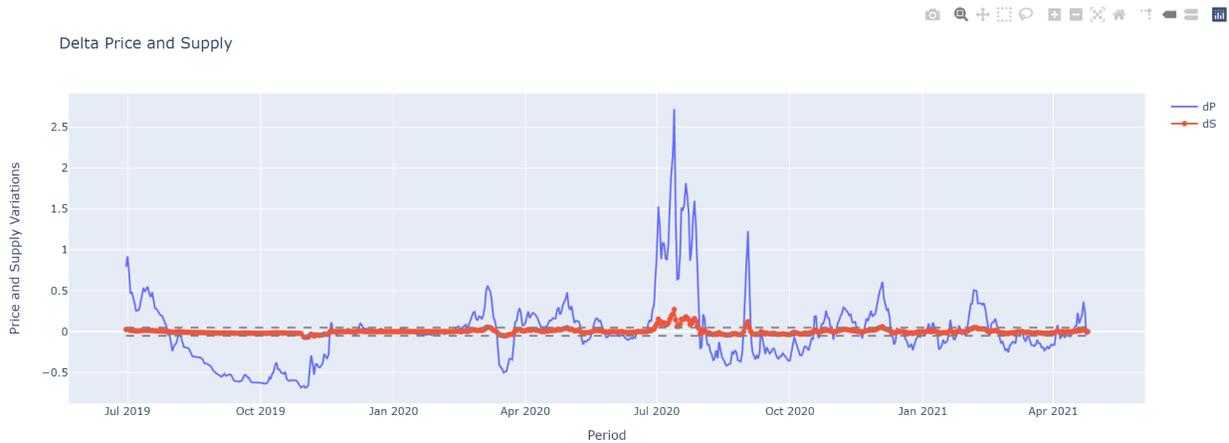

Figure 9: the deviation of AMPL price from its target and the supply volatility of AMPL.

Unlike other types of stablecoins, algorithmic stablecoins maintain price stability without relying on external collaterals such as fiat or other cryptocurrencies (Sidorenko 2019, Moin et al. 2020, Lyons and Viswanath-Natraj 2020). Nevertheless, the collateral-free stability mechanism is a double-edged sword. On the one hand, it allows policymakers to achieve desirable outcomes by selecting the optimal policy parameter, which provides a straightforward toolbox for not only the private sector stablecoin design but also for governments and central bankers in designing and effectively implementing central bank digital currencies (CBDCs) (Dell'Erba 2019). On the other hand, price stability is susceptible to market sentiment. For instance, when price depreciates, users and investors might flee out of panic, which further aggravates price fluctuations. (Avernas, Bourany, and Vandeweyer 2021). Therefore, designing a more resilient monetary policy is vital for the sustainability of algorithmic stablecoins.

---

[12] A =0.05, B start with B=30 and change to B=10 as of 2019/10/30.